\documentclass[prl,aps,showpacs,nofootinbib,twocolumn]{revtex4}
\usepackage{epsfig}
\usepackage{amssymb}
\usepackage{amsmath}
\usepackage{amsfonts}
\def\rhob{{\boldsymbol\rho}}
\begin{document}
\title{Heavy deformed nuclei in the shell model Monte Carlo method}
\author{Y. Alhassid$^1$, L. Fang$^1$, and H. Nakada$^2$}
\affiliation{$^{1}$Center for Theoretical Physics, Sloane Physics
Laboratory,\\ Yale University, New Haven, CT 06520, USA\\
$^{2}$Department of Physics, Graduate School of Science,\\
Chiba University, Inage, Chiba 263-8522, Japan
}
\def\be{\begin{equation}}
\def\ee{\end{equation}}
\begin{abstract}
We extend the shell model Monte Carlo approach to heavy deformed nuclei using a new proton-neutron formalism. The low excitation energies of such nuclei necessitate calculations at low temperatures for which a stabilization method is implemented in the canonical ensemble. We apply the method to study a well deformed rare-earth nucleus, $^{162}$Dy. The single-particle model space includes the  $50-82$ shell plus $1f_{7/2}$ orbital for protons and the $82-126$ shell plus $0h_{11/2}, 1g_{9/2}$ orbitals for neutrons. We show that the spherical shell model reproduces well the rotational character of $^{162}$Dy within this model space. We also calculate the level density of $^{162}$Dy and find it to be in excellent agreement with the experimental level density, which we extract from several experiments.
\end{abstract}
\maketitle

{\em Introduction.}
The shell model Monte Carlo (SMMC) approach \cite{la93,al94} has been successful in the calculation of statistical nuclear properties such as partition functions and level densities \cite{NA97,ALN99}. However, most of the SMMC calculations carried out to date have been limited to medium-mass nuclei whose deformation is not particularly large and whose low-temperature properties are well described by a single major shell. In such even-even nuclei, the gap to the first excited state is $\sim 1-2$ MeV and the ground state can be reached in practice with moderate values of the inverse temperature  $\beta\sim 3$ MeV$^{-1}$.

For the SMMC to be useful across the table of nuclei, it is neessary to show its applicability in heavy nuclei, e.g., the rare-earth region, where the deformation in mid-shell nuclei can be large and the first excitation energy is $\sim 100$ keV. Such nuclei present a difficult technical challenge in SMMC since it is necessary to propagate to much large values of $\beta \sim 20$ MeV$^{-1}$. At moderate and large values of $\beta$, the propagator becomes ill-conditioned  and one must  stabilize the propagation, keeping its large and small scales separated. Stabilization methods were developed in strongly correlated electron systems in the grand-canonical ensemble~\cite{LG92}. However, nuclear applications require use of the canonical ensemble, for which stabilization methods are considerably slower. An important issue is whether it is possible to describe the known rotational behavior of strongly deformed nuclei in the framework of a truncated spherical shell model. Here we provide an affirmative answer, demonstrating our methods for the well-deformed nucleus $^{162}$Dy. This is the largest SMMC calculation to date.

{\em SMMC in proton-neutron formalism.} Since protons and neutrons occupy different shells, the isospin formalism is no longer valid and it is necessary to recast SMMC in a proton-neutron formalism. A formulation based on $T_z$ projection was used in Ref.~\cite{oz06}. Here we use a more efficient formulation in which protons and neutrons are treated explicitly. A single-particle orbital $a$ has good quantum numbers $n,l,j$  and is $2j+1$-fold degenerate (in magnetic quantum number $m$) with energy $\epsilon_a$. We assume that the single-particle model space includes $N_s^p$ orbitals for protons (including the magnetic degeneracy) and $N_s^n$ orbitals for neutrons.  The two-body interaction matrix elements are given by $V^{pp}_J$,  $V^{nn}_J$ and $V^{pn}_J$ for proton-proton, neutron-neutron and proton-neutron, respectively. We first rewrite the two-body interaction in a density decomposition by performing a Pandya transformation for each type of matrix elements to obtain the matrices $E^{pp}_K$, $E^{nn}_K$ and $E^{pn}_K$. Defining the matrix ${\bf E}_K$ as the $2\times 2$ block structure with $E^{pp}_K$ and $E^{nn}_K$ as the diagonal blocks and $E^{pn}_K$, $E^{np}_K =\left(E^{pn }_K\right)^T $  as the off-diagonal blocks, we have
\begin{equation}\label{Hamiltonian}
\!\! H \!\! = \!\!\sum_a \epsilon'_a \hat n_a + \sum_r \epsilon'_r \hat n_r + \frac{1}{2} \sum_{K M} (-)^M \rhob^T_{K -M} {\bf E}_K \rhob_{K M} \;,
\end{equation}
where the column vector $\rhob_{KM}$ is composed of both proton and neutron densities, and $\epsilon'_a$ ($\epsilon'_r$) are shifted proton (neutron) single-particle energies (the shift originates in the Pandya transformation).
The matrix ${\bf E}_K$ is real symmetric and can be diagonalized by an orthogonal transformation. The quadratic two-body term in (\ref{Hamiltonian}) can then be written as $H_2'= \frac{1}{2}\sum_{K \alpha} \lambda_{K \alpha} \sum_M (-)^M \rho_{K -M}(\alpha) \rho_{KM}(\alpha)$, with $\lambda_{K\alpha}$ being the eigenvalues of ${\bf E}_K$. The eigenvectors $\rho_{KM}(\alpha)$ are linear combinations of proton and neutron densities.

In the Condon-Shortely convention, the time-reversed density is given by $\bar \rho_{KM}(ac) = \tilde\pi (-)^{K+M} \rho_{K -M}(ac)$, where $\tilde\pi \equiv (-)^{l_a+l_c}$ is the particle-hole parity. We can then rewrite the two-body part of the Hamiltonian as $H'_2
 =  \frac{1}{2} \sum_{K \alpha} V_{K \alpha} \sum_M \left[ Q_{KM}^2(\alpha)+  R_{KM}^2(\alpha)\right]$,
where $V_{K \alpha} = \tilde\pi (-)^{K} \lambda_{K\alpha}$, and $Q_{KM}$, $R_{KM}$ are the real and imaginary parts of $\rho_{KM}$ (where complex conjugation is defined by time reversal).  A Hubbard-Stratonovich (HS) transformation can be directly applied to this quadratic form. The resulting decomposition has a good Monte Carlo sign when $V_{K \alpha} <0$ for all $K, \alpha$.  The one-body Hamiltonian of the propagator $U_\sigma$ in the HS integrand is a linear combination of proton and neutron densities and the corresponding propagator is a product of a proton and a neutron one-body propagators. The computational cost is thus smaller than the method used in Ref.~\cite{oz06}, in which the dimension of the propagator matrix is $N_s^p+N_s^n$.

{\em Stabilization.}
In SMMC, the evolution operator for a given sample is calculated as a product of one-body propagators of time slice $\Delta \beta$. The number of matrix multiplications increases with $\beta$ and the propagator matrix might become ill-conditioned, i.e., the ratio of its largest to smallest eigenvalue is too large. Large and small numerical scales get mixed in the propagation, resulting in the loss of important information.

A method was proposed to stabilize matrix multiplication in the grand-canonical formulation \cite{LG92}. The method is based on the decomposition of a matrix $M$ into the form $M=ADB$ where $A,B$ are well-behaved under multiplication and
$D$ is a diagonal matrix whose elements are positive numbers
containing the various scales. In the singular
value decomposition (SVD), the matrix $M$ has the form $M=UDV$ where $U$ and
$V$ are unitary matrices. In the modified Gram-Schmidt (MGS) decomposition $M=LDV$ or $M=UDR$ where $L$ ($R$) is a lower (upper) triangular matrix with diagonal elements $1$.  We have adopted the MGS decomposition, which can be up to $\sim 20$ times faster than SVD \cite{go96}. In SMMC, it is necessary to stabilize the canonical propagator. Since the canonical formulation is accomplished by a particle-number projection, each term in the quadrature sum must be stabilized.

{\em Choice of model space and interaction.}
To describe the rotational character of a mid-shell rare-earth nucleus, it is necessary to use a sufficiently large single-particle model space.
To determine the required single-particle orbitals, we consider a Woods-Saxon (WS) plus spin-orbit mean-field potential. The spherical orbitals of this potential are  $|\alpha jm\rangle$ ($\alpha$ represents the remaining quantum numbers). Introducing an axial deformation $\beta_2$ in the WS potential, we determine its eigenstates $|k m\rangle$ and expand them in the spherical orbitals,
$|km\rangle = \sum_{\alpha j} c^m_{k;\alpha j}|\alpha jm\rangle$. The spherical occupations are then given by
$r_{\alpha j} = \frac{1}{2j+1}\sum_{km} |c^m_{k;\alpha j}|^2
\langle n_{km}\rangle$,
where $\langle n_{km}\rangle$ are the occupations of the deformed orbitals (1 below the Fermi energy and 0 above). In a shell model approach,  we should include in our model space the physically important spherical orbitals, while the influence of all other orbitals is taken into account by renormalizing the interaction. Here we include the orbitals that satisfy $0.1<r_{\alpha j}<0.9$ at $\beta_2=0.35$. This determined the model space to be
$0g_{7/2}$, $1d_{5/2}$, $1d_{3/2}$, $2s_{1/2}$, $0h_{11/2}$, $1f_{7/2}$
for protons, and $0h_{11/2}$, $0h_{9/2}$, $1f_{7/2}$, $1f_{5/2}$, $2p_{3/2}$, $2p_{1/2}$, $0i_{13/2}$, $1g_{9/2}$ for neutrons. This model space includes orbitals outside the corresponding $0 \hbar\omega$ major shells,
in contrast to Ref.~\cite{wkd00}.

As an effective interaction, we use the dominant collective parts of realistic interactions: monopole pairing and multipole-multipole interactions. This interaction is similar to the one used in Ref.~\onlinecite{NA97} except that protons and neutrons occupy different orbitals
\begin{equation}
\!\! - \!\!\!\!\sum_{\nu=p,n} g_\nu P^\dagger_\nu P_\nu
 - \!\!\sum_\lambda \chi_\lambda :(O_{\lambda;p} + O_{\lambda;n})\cdot
 (O_{\lambda;p} + O_{\lambda;n}): \;.
\label{SMint}\end{equation}
Here $P^\dagger_\nu = \sum_{nljm}(-)^{j+m+l}a^\dagger_{\alpha jm;\nu} a^\dagger_{\alpha j-m;\nu}$ ($\nu=p,n$) is the $J=0$ pair creation operator, $::$ denotes normal ordering, and $O_{\lambda;\nu}=\frac{1}{\sqrt{2\lambda+1}}
\sum_{ab}\langle j_a||\frac{d V_{\rm WS}}{dr}Y_\lambda||j_b\rangle
[a^\dagger_{\alpha j_a;\nu}\times \tilde{a}_{\alpha j_b;\nu}]^{(\lambda)}$
is a surface-peaked multipole operator [$\tilde{a}_{jm}=(-1)^{j-m}a_{j-m}$].
We include quadrupole, octupole and hexadecupole terms
(\textit{i.e.}, $\lambda=2,3,4$) with corresponding strengths
 $\chi_\lambda=\chi\cdot k_\lambda$. The parameter $\chi$ is determined self-consistently~\cite{abdk96} and
$k_\lambda$ are renormalization factors accounting for core polarization effects.

To determine $k_2$, we note that the ``slope" of $\ln \rho(E_x)$ ($\rho(E_x)$ is the total level density) at higher energies is sensitive to $\chi_2$. We find that a value of $k_2=2.12$ reproduces the slope of the experimental $\ln \rho(E_x)$ in the finite-temperature Hartree-Fock-Bogolyubov (HFB) approximation.
This value is close to the value of $k_2=2$ used in Ref.~\onlinecite{NA97}.
For the octupole and hexadecupole interactions we take $k_3=1.5$ and $k_4=1$ \cite{NA97}.

In Ref.~\onlinecite{NA97} we determined the pairing strength to reproduce the experimental odd-even mass differences in neighboring spherical nuclei using  number-projected BCS calculations. Following a similar method
for spherical nuclei in the mass region $Z=50-82$, $N=82-126$, we obtain $g_p=10.9\,\mathrm{MeV}/Z$ ($g_n=10.9\,\mathrm{MeV}/N$). Here we find however that a reduction in the value of $g_p$ and $g_n$ is necessary to reproduce the moment of inertia of the ground-state band, and use a reduction factor of 0.77 (see below). Part of this reduction may be ascribed to fluctuations of the pairing fields.

For the one-body Hamiltonian we use the single-particle orbitals of the spherical WS plus spin-orbit potential. Since the WS potential represents a mean-field potential, we determine the bare single-particle energies so they reproduce the WS single-particle energies in the Hartree-Fock (HF) approximation.

\begin{figure}[t]
  \epsfxsize= 0.9\columnwidth \centerline{\epsffile{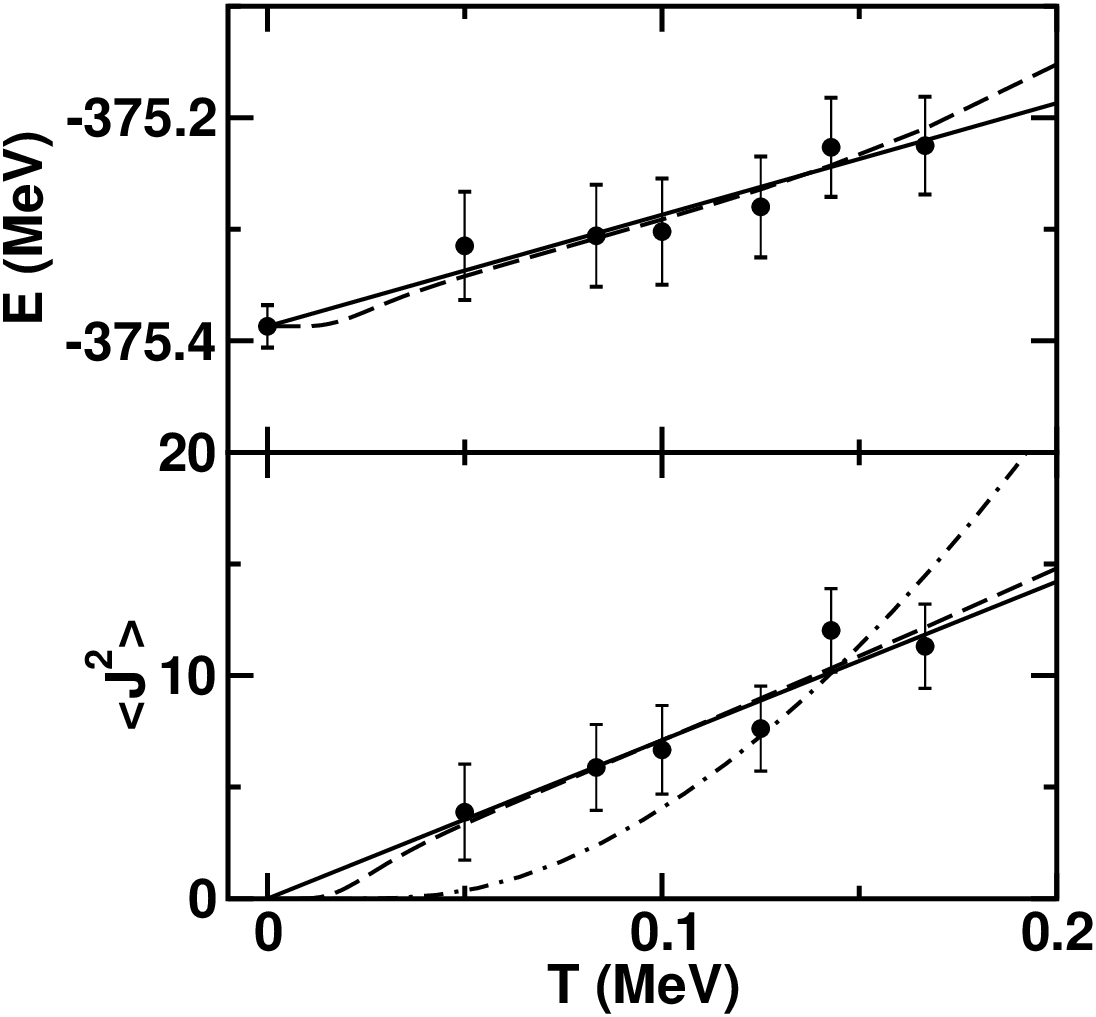}}
\caption{ Low-temperature thermal energy $E$ (top panel) and $\langle \bf J^2\rangle$ (bottom panel) versus temperature $T$ in $^{162}$Dy. The SMMC results (solid circles)  are fitted to (\ref{band}) (solid lines). The dashed lines are results obtained from the lowest five experimental bands in $^{162}$Dy. The dashed-dotted line (bottom panel) is a fit to the vibrational model result.
}
\label{low-T}
\end{figure}

{\em Ground-state energy and moment of inertia}. We demonstrate our methods for a typical strongly deformed rare-earth nucleus, $^{162}$Dy.
To determine the ground-state energy it is necessary to extrapolate the thermal energy to $\beta = \infty$. We carried out stabilized calculations at large $\beta$ values (up to $\beta=20$ MeV$^{-1}$) using time slices of $\Delta \beta=1/32$ MeV$^{-1}$  and $\Delta \beta=1/64$ MeV$^{-1}$. The SMMC thermal energy is shown versus $T$ in the top panel of Fig.~\ref{low-T} (solid circles).  The bottom panel of Fig.~\ref{low-T} shows $\langle {\bf J}^2\rangle$ versus $T$ (${\bf J}$ is the total angular momentum).

A simple way to extract the ground-state energy $E_0$ is to assume a ground-state rotational band $E_J = E_0 + \hbar^2 J(J+1)/2{\cal I}_g$ with a moment of inertia ${\cal I}_g$. At sufficiently low temperatures, only the ground band contributes to thermal observables and a simple calculation gives
\begin{equation}\label{band}
E(T) \approx E_0 + T\;,\;\;\;\; \langle {\bf J}^2 \rangle\approx 2{\cal I}_g T/\hbar^2 \;.
\end{equation}

Fitting a straight line of slope 1 to $E(T)$, we find $E_0=-375.387\pm 0.019$ MeV. Comparing with the HFB ground-state energy of $E_{\rm HFB}= -372.263$ MeV, we determine a correlation energy of $E_{\rm HFB} - E_0 =3.124 \pm 0.019 $ MeV.

By fitting a straight line $2{\cal I}_g T/\hbar^2$ to $\langle {\bf J}^2\rangle$, we also determine the moment of inertia ${\cal I}_g/\hbar^2 =35.8\pm 1.5$ MeV$^{-1}$ of the ground-state band. This value agrees with the experimental value of $ 37.2$ MeV$^{-1}$, extracted from the excitation energy of the first $2^+$ state ($80.7$ keV).

The SMMC results for $\langle {\bf J}^2 \rangle$ agree with the second relation in (\ref{band}), derived under the assumption of a rotational band. This provides evidence that our model space is sufficient to reproduce the rotational behavior of this strongly deformed nucleus within a truncated shell model approach.  We also show in Fig.~\ref{low-T} results of a fit to $\langle {\bf J}^2 \rangle$  assuming a vibrational model (dotted-dashed line). Our SMMC results clearly indicate that the low-lying levels of our shell model Hamiltonian are not vibrational.

To test the validity of the one-band approximation, we show in Fig.~\ref{low-T} results for $E(T)$ and $\langle {\bf J}^2 \rangle$ calculated using the five lowest experimental bands in  $^{162}$Dy (dashed lines). The one-band expressions (\ref{band}) are seen to be valid for $T \lesssim 0.16$ MeV down to $T \approx 0.05$ MeV.

\begin{figure}[t]
  \epsfxsize= 0.85 \columnwidth \centerline{\epsffile{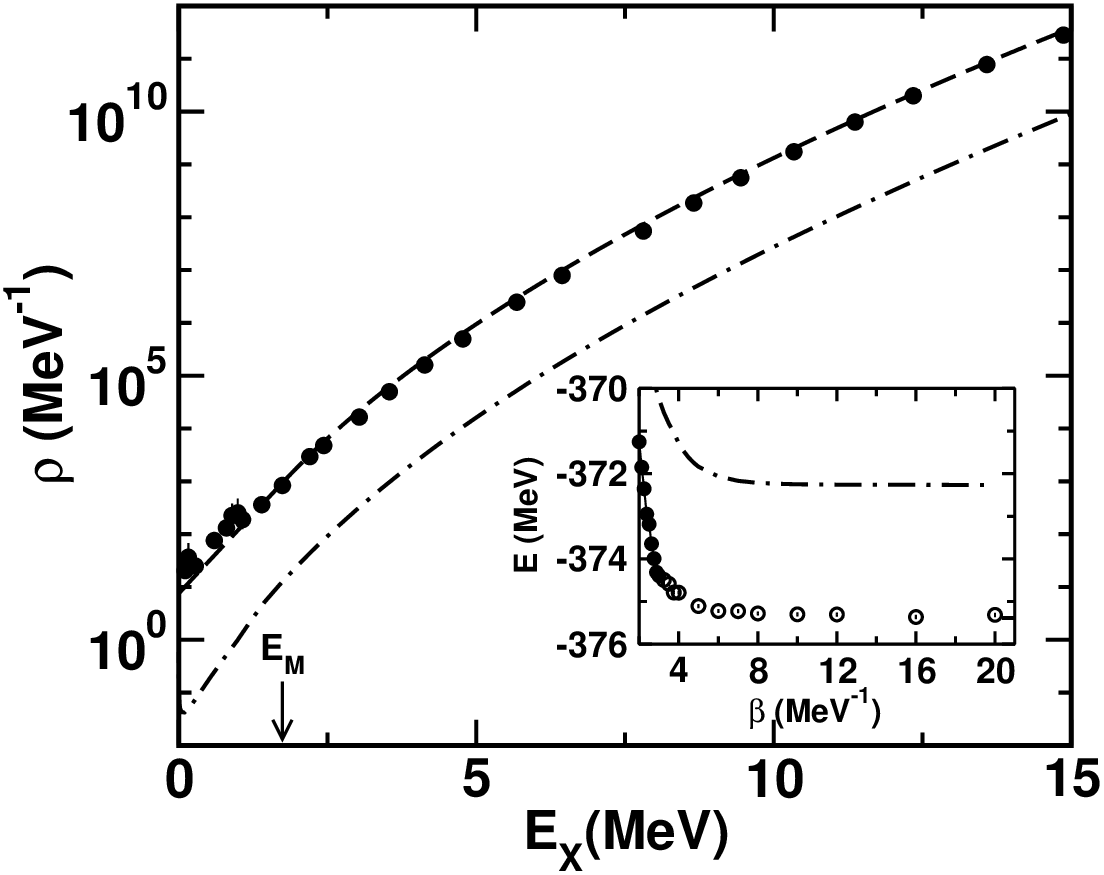}}
\caption{ Total state density of $^{162}$Dy. The SMMC results (solid circles) are compared with the experimental level density (dashed line) and the HFB level density (dotted-dashed line). Inset: thermal energy E versus $\beta$ (for $\beta >2$ MeV$^{-1}$) in SMMC (circles) and in HFB (dotted-dashed line). The open circles indicate stabilized calculations.
}
\label{level-smmc}
\end{figure}

{\em Level density}.  We use the saddle-point expression for the level density in terms of the canonical entropy and heat capacity~\cite{NA97}, which in turn can be extracted from the thermal energy $E(\beta)$. Discretization of $\beta$ introduces systematic errors in $E(\beta)$ and we found it necessary to extrapolate to $\Delta \beta=0$ using two time slices $\Delta \beta=1/32, 1/64$ MeV$^{-1}$. For $\beta \leq 3.25$ MeV$^{-1}$ we used a linear extrapolation in $\Delta \beta$ while for larger values of $\beta$ the dependence on $\Delta \beta$ is weaker and we took an average value. The results for $E(\beta)$ (stabilized for $\beta > 3$ MeV$^{-1}$)
 are shown in the inset of Fig.~\ref{level-smmc}. For comparison we  also show $E(\beta)$ in the HFB approximation (dotted-dashed line).

The SMMC level density is shown by the solid circles in Fig.~\ref{level-smmc}. It agrees very well with a composite-formula level density (dashed line), which we extract from several experiments (see below). For comparison, we also show the HFB level density (dotted-dashed line). We observe strong enhancement of the SMMC level density relative to the HFB density. Indeed the latter describes only the intrinsic states while the SMMC results include all states and in particular the rotational bands.

{\em Experimental Level Density.}
There are various experimental data that can be used to determine the level density of $^{162}$Dy : an almost complete level scheme at low excitations ($E_x \lesssim 2$ MeV) \cite{isotope,ap06,rick}, neutron resonance data at $E_x=8.196$ MeV \cite{iaea98} and data obtained by the so-called Oslo method~\cite{oslo,gu03}.

In our level density studies in mid-mass nuclei, we used a back-shifted Bethe formula (BBF) to parametrize the SMMC level density and compared with similarly parametrized data \cite{dilg73}. Here we find that at low excitations a constant temperature formula works better than the BBF.
We therefore use a composite formula~\cite{gc65} that combines a constant temperature formula and a BBF

\be\label{composite} \rho(E_x)=\left\{
\begin{array}{l} {\displaystyle \exp \left[(E_x-E_1)/ T_1\right]
\;\;\;\;\;\;\;\;\;\;\;E_x<E_M }\\
 {\pi^{1/2}
a^{-1/4} \over 12 (E_x-\Delta)^{5/4}}e^{2\sqrt{a(E_x-\Delta)}}\;\;\;\;
E_x>E_M\;.
\end{array} \right.
\ee
The two formulas are matched at an energy $E_M$ assuming the continuity of the level density and its derivative.

\begin{figure}[t!]
  \epsfxsize= 0.9 \columnwidth \centerline{\epsffile{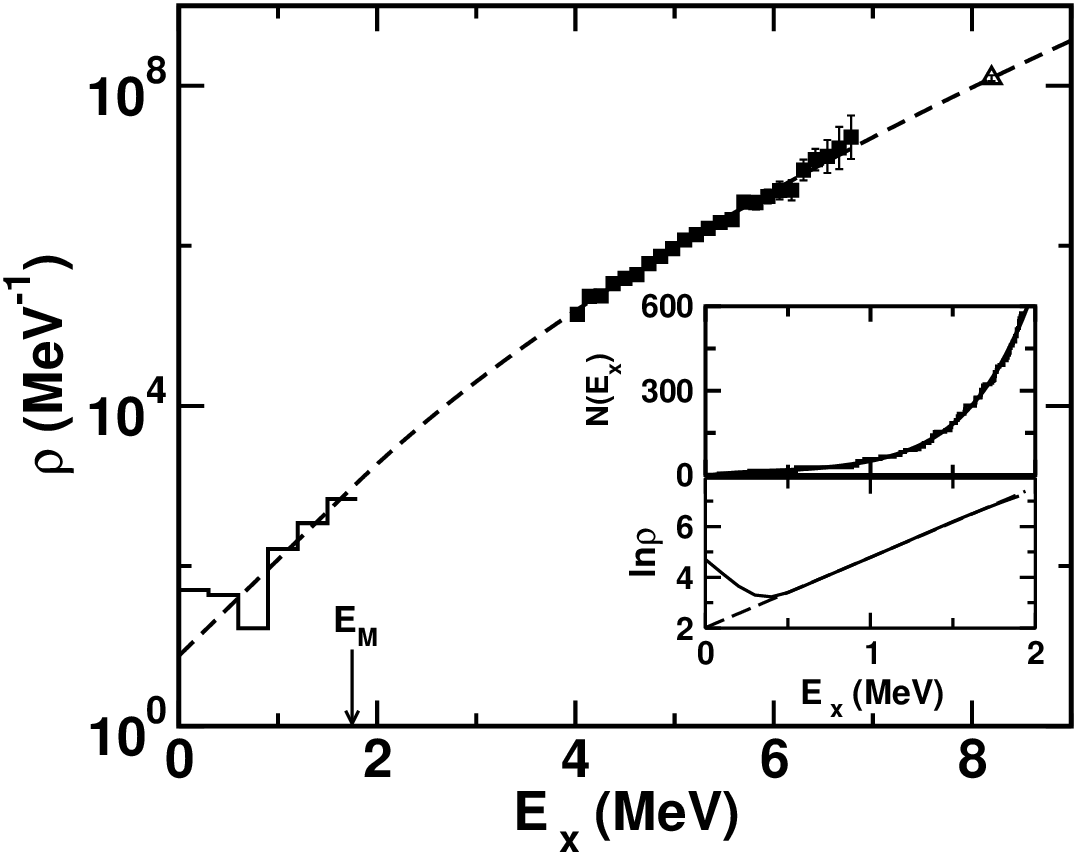}}
\caption{ Experimental state density of $^{162}$Dy. Histograms are from level counting~\cite{isotope,ap06,rick}, solid squares are renormalized Oslo data \cite{gu03} and the triangle is the neutron resonance data \cite{iaea98}. The dashed line is a fit to the composite formula (\ref{composite}). Inset (top): the experimental staircase function $N(E_x)$ (histograms) and its fit to a sixth-order polynomial (solid line). Inset (bottom): average level density obtained by a derivative of the fit to $N(E_x)$ (solid line) and its fit to a constant temperature formula (dashed line).
}
\label{exp-density}
\end{figure}

To determine the state density at low energies we construct the
 staircase function $N(E_x)$ (counting number of states below $E_x$) and fit it to a sixth order polynomial (top panel of inset to Fig.~\ref{exp-density}). Its derivative (solid line in botom panel of inset) describes the average state density $\rho(E_x)$.
We observe that $\ln\rho(E_x)$ is well fitted by a straight line in the
 range $0.6 < E_x < 1.8$ MeV,  determining  $E_1=-0.73$
 MeV and $T_1=0.36$ MeV. For a given $E_M$, the parameters $a$ and $\Delta$ in (\ref{composite}) are determined by the matching conditions.
$s$-wave neutron resonance data determines the sum of the level
 densities for spin $I\pm 1/2$ ($I$ is the spin of the target nucleus) at
 the neutron separation energy.
Using a spin cutoff parameter of $\sigma^2={\cal I}T / \hbar^2$,
 with the rigid-body moment of inertia ${\cal I} \approx 0.015
 A^{5/3}\hbar^2$ and $T=\left[(E_x-\Delta)/ a\right]^{1/2}$,
 we obtain the total level density at the neutron resonance energy $E_x = 8.196$ MeV, where the spin cutoff model is valid~\cite{al06}.
Additional data are from recent gamma-ray spectroscopy experiments
 \cite{gu03} using the Oslo method \cite{oslo},
which determines $\ln\rho(E_x)$ up to $b_0+b_1 E_x$ with appropriate constants $b_0$ and $b_1$. The measured level density is converted to state density using
a rigid-body moment of inertia. Since at low excitations the moment of inertia is reduced from its rigid-body value, we only use data for $E_x \geq 4$ MeV. The matching energy $E_M$, together with $b_0$ and $b_1$,
are determined by a $\chi^2$ fit of $\ln\rho(E_x)$ to the Oslo and neutron resonance data. We find  $E_M= 1.752 \pm 0.036$ MeV, which in turn determines $a=18.28 \pm 0.15$ MeV$^{-1}$ and $\Delta = 0.421 \pm 0.014$ MeV. The corresponding composite density is shown by the dashed lines in Figs.~\ref{exp-density} and \ref{level-smmc}, and is in very good agreement with the SMMC state density.

{\em Conclusion.}
We have extended the shell model Monte Carlo approach to heavy deformed nuclei
using a new proton-neutron formalism. A stabilization method is implemented in the canonical ensemble to accurately describe the low-energy properties of such nuclei. Applying the method to $^{162}$Dy, we show that the spherical shell model approach reproduces well the rotational character of this nucleus, as long as a sufficiently large model space and an appropriate effective Hamiltonian are used. We also calculate the level density of $^{162}$Dy and find it to be in excellent agreement with the experimental level density, extracted from several experiments.

This work is supported in part by the U.S. DOE grant No. DE-FG-0291-ER-40608
and as Grant-in-Aid for Scientific Research (B), No.~15340070, by the MEXT,
Japan. Computational cycles were provided by the NERSC high performance computing facility at LBL.

\end {document}